# Science Requirements and Performances for EAGLE for the E-ELT

C. J. Evans[1,*], M. D. Lehnert[2], J.-G. Cuby[3], S. L. Morris[4], M. Puech[2],
N. Welikala[3], A. M. Swinbank[4] & H. Schnetler[1]

[1] UK Astronomy Technology Centre, Royal Observatory Edinburgh, Blackford Hill, Edinburgh, EH9 3HJ, UK
[2] GEPI, Observatoire de Paris, 5 Place Jules Janssen, 92195 Meudon Cedex, France
[3] Laboratoire d'Astrophysique de Marseille, BP8, 13376 Marseille Cedex 12, France
[4] Department of Physics, Durham University, South Road, Durham, DH1 3LE, UK

**ABSTRACT**

EAGLE is a Phase A study of a multi-IFU, near-IR spectrometer for the European Extremely Large Telescope (E-ELT). The design employs wide-field adaptive optics to deliver excellent image quality across a large (38.5 arcmin$^2$) field. When combined with the light grasp of the E-ELT, EAGLE will be a unique and efficient facility for spatially-resolved, spectroscopic surveys of high-redshift galaxies and resolved stellar populations. Following a brief overview of the science case, here we summarise the functional and performance requirements that flow-down from it, provide illustrative performances from simulated observations, and highlight the strong synergies with the *James Webb Space Telescope (JWST)* and the Atacama Large Millimeter Array (ALMA).

## 1. INTRODUCTION

One of the dominant scientific motivations for the next generation of ground-based telescopes is the study of galaxy evolution. This encompasses a huge range of phenomena and scales – from studies of resolved stars in nearby galaxies, to our attempts to understand the properties of the most distant galaxies in the very young Universe. From consideration of these cases, and the broader astrophysical landscape, a set of science requirements were compiled for a near-infrared spectrograph with multiple, deployable integral-field units (IFUs). These requirements were used to develop the EAGLE concept (Cuby et al., these proceedings), which combines the light-gathering power of the E-ELT with excellent image quality, via adaptive optics (AO) correction, across a wide field-of-view.

An overview of the principal EAGLE science cases was given recently by Evans et al. (2010a). In brief, these comprise:

- **Physics of high-redshift galaxies:** Models of galaxy formation and evolution rely heavily on analytic prescriptions with inputs from existing observations such as metallicity, angular momentum, the stellar initial mass function, the spatial distribution of the gas, and the frequency of major/minor mergers. These models can only be tested rigorously with a sufficiently large sample (1000s of galaxies) and over a large enough volume to rule out field-to-field biases ('cosmic variance').

  Until recently, the study of high-redshift galaxies was generally limited to redshift determinations and analysis of the integrated spectrum from each system. The availability of IFUs on 8-10 m class telescopes has heralded a new era of galaxy studies, where we can derive *spatially-resolved* kinematics and physical properties of selected distant galaxies up to $z \sim 3$ (e.g. Förster Schreiber et al. 2006, 2009; Flores et al. 2006; Puech et al. 2006; Yang et al. 2008; Lehnert et al. 2009). However, these are typically only the most massive, high surface-brightness systems. The E-ELT offers the potential for spatially-resolved observations of an unbiased and unprecedented sample of high-redshift galaxies. This will provide us with a clear view of galaxy evolution over a broad range of redshifts and galaxy types for the first time.

- **'First light': Characterizing the most distant galaxies:** Observational constraints on the properties of galaxies at $z > 6$ are scarce, with only a handful of galaxies confirmed at this epoch. Gaining an inventory of the basic properties of the first galaxies (at $7 < z < \sim20$) is one of the greatest challenges in modern astronomy. If we can understand the formation of the first stars, assembly of the first galaxies, and the growth of super-massive black holes through gas accretion, we can build a complete picture of the star formation and quasar activity thought to be responsible for the reionization of the Universe.

---

[*] chris.evans@stfc.ac.uk

The high-redshift ($z \geq 7$) objects recently discovered by the *HST*-WFC3 (Oesch et al., 2010; Bouwens et al., 2010; McLure et al. 2010; Fig. 1) are merely a taste of what awaits in the coming decade. Further deep imaging with the *HST*, ground-based telescopes (e.g. VISTA, VLTs) and, ultimately, the *JWST*, will discover large samples of very distant Lyman-break and Lyman-$\alpha$ emitting galaxies. One of the key goals for the E-ELT is to probe their properties. For example, some of the objects recently discovered with WFC3 are likely beyond the spectroscopic sensitivity of the *JWST*. Discovery is merely the first step – understanding their physical properties is an entirely different matter, requiring follow-up spectroscopy with an ELT.

- **Resolved stellar populations in the Local Volume:** Recent discoveries of disrupted satellite galaxies and extended outer-galaxy halos have shown that our evolutionary picture of the Milky Way and other Local Group galaxies is far from complete. Resolved stellar populations provide us with a valuable fossil record of the past star formation and interaction histories of these galaxies. While photometric studies are immensely powerful, only when armed with chemical abundances and stellar kinematics can we break the age-metallicity degeneracy and also disentangle the populations associated with different structures, i.e. follow-up spectroscopy is required.

  The sensitivity of the E-ELT, combined with excellent angular resolution (via AO), will bring spectroscopic studies of individual evolved stars at Mpc distances within our grasp for the first time. This will unlock a wealth of new and exciting targets in which we can study galaxy evolution directly. This 'stellar archaeology' will provide unique inputs toward future theoretical efforts regarding galaxy formation and evolution. Only by looking beyond the Local Group can we investigate the histories of a broad range of galaxies, including other spiral-dominated groups, starbursts, the nearest ellipticals, and very metal-poor irregulars.

These cases are used to derive instrument requirements in the next section. In the course of quantifying these, detailed science simulations were undertaken, examples are which are shown in Section 3, to provide illustrative performances.

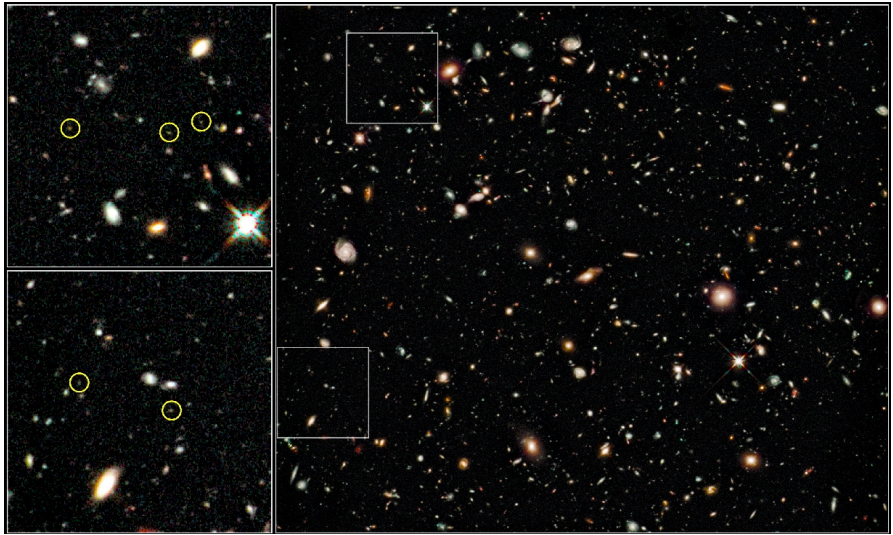

**Fig. 1:** High redshift ($z \geq 7$) candidates in the *HST*-WFC3 Ultra Deep Field (NASA/ESA/Illingworth/Bouwens). Discovery of large samples of such galaxies is a cornerstone of the *JWST* mission – EAGLE will provide the detailed follow-up spectroscopy.

## 2. REQUIREMENTS FLOW-DOWN

The EAGLE instrument requirements are now discussed, in reference to the cases which are the strongest drivers. Note that in the Phase A study we have assumed that EAGLE will be mounted at the planned Gravity Invariant Focal Station (GIFS) of the E-ELT, in which the instrument is located at Nasmyth, but with the focal-plane parallel to the ground (avoiding a variable gravity vector on the instrument with time).

## 2.1. Patrol Field & Multiplex

The requirements which dominate the scale of the instrument are the total field-of-view on the sky from which targets can be selected (the 'patrol field') and the number of objects/fields to be observed simultaneously (the 'multiplex'). Source densities of the relevant high-redshift galaxies are of order of 1 to 2 arcmin$^{-2}$ (e.g., Förster-Schreiber et al. 2004; Marchesini et al. 2007; Reddy et al. 2008), reducing to ~0.5 arcmin$^{-2}$ for the highest redshift targets at $z > 7$ (e.g., Fig. 4 of Bouwens et al. 2010, to $M_{AB}$ ~ 28). With relatively low target densities such as these, one has to consider the potential field-of-view delivered by the telescope before placing requirements on the instrument design. The notional focal plane at the GIFS is 5 arcmin (diameter), but a larger annulus, with an external diameter of 10 arcmin, is used by the E-ELT for wavefront sensing and other functions. The telescope has been designed for good image quality in this region, so if one contemplates efficient surveys of high-redshift galaxies (in which the number of 'good' sources will be lower than published source densities owing to the night-sky lines), a compelling requirement is to access as much of the potential field as possible. In the EAGLE design we have therefore sought to provide as large a patrol field as possible, within the constraints of the required interfaces to the telescope and the optics necessary to select and monitor the telescope laser guide stars for AO. This approach is also warranted in the broader context of EAGLE as a facility-class instrument to address a wide range of science cases and targets, i.e., maximising its flexibility and capability.

To illustrate this point in more detail, we estimated the number counts of emission-line galaxies in the EAGLE field-of-view provided by the Phase A design (with an equivalent on-sky area of 38.5 arcmin$^2$). To study the mass assembly and physics of galaxies between $1 < z < 4$, we ideally want to trace mass-selected samples of emission-line galaxies in different redshift intervals. In the absence of sufficiently deep spectroscopic surveys from which to select a homogeneous sample of galaxies, we estimated the source counts using photometric redshift surveys. Using results from the COSMOS survey (Scoville et al. 2007), we selected a blue population of galaxies with $M_{AB}(NUV) - M_{AB}(r') < 3.5$, to approximate the emission-line population to $I_{AB} \leq 26$. The sample is complete to $z \sim 4$ and $M_U < -19$. As a first-order estimate of the number of potential EAGLE targets, we selected star-forming galaxies on the basis of $M_U$ (as a proxy for the star-formation activity) for different redshift intervals over $1 < z < 4$. The left-hand panel of Fig. 2 shows the number counts of galaxies in the two main intervals of $M_U$, comprising a population with $-22 < M_U < -21$ (red panels) and a fainter population with $-21 < M_U < -20$ (blue).

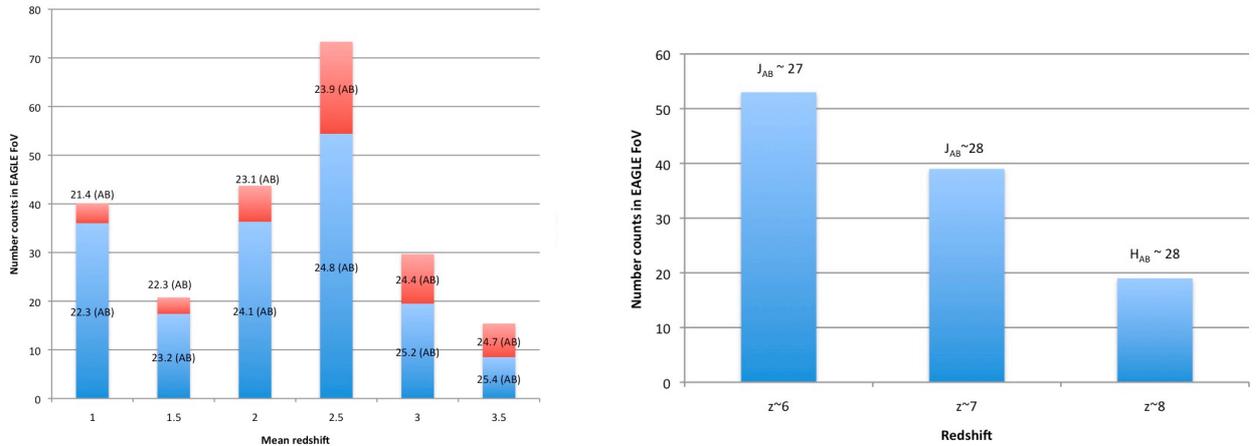

**Fig. 2:** *Left-hand panel:* Target densities in the 38.5 arcmin$^2$ EAGLE patrol field for galaxies at $1 < z < 4$; abscissae are lower limits of intervals with $\Delta z = 0.5$. Bright systems ($-22.0 < M_U < -21.0$) are in red, fainter systems ($-21.0 < M_U < -20.0$) are in blue. *Right-hand panel:* Target densities for galaxies at $z \geq 6$.

We also folded-in the spectroscopic success rate of obtaining a redshift from any of the Hα, [OII] λ3729Å, or [OIII] λλ4959, 5007Å emission lines into the final target counts. Fig. 2 is therefore an estimate of the number of star-forming galaxies at each redshift with measurable line characteristics (i.e. where the diagnostic lines avoid the strong night sky lines). We thus expect to spectroscopically detect ~40 galaxies at $z \sim 1$ (mean $J_{AB}$ ~21.4), with 70 galaxies at $z \sim 2.5$ (mean $J_{AB}$ ~ 23.9), and 30 by $z \sim 3$ (mean $J_{AB}$ ~ 24.4).

In the right-hand panel of Fig. 2 we show the corresponding number counts for galaxies at $6 < z < 8$, estimated from the surface densities of (a) $i_{775}$ dropouts at $z \sim 6$ in the *HST* ultra-deep field (Bouwens et al. 2008); (b) $z_{850}$ dropouts at $z \sim 7$

(Oesch et al. 2009); (c) $Y_{105}$ dropouts at $z \sim 8$ (Bouwens et al. 2010). The spectroscopic success rates for rest-frame UV metal lines were then folded into the calculation, to produce the final target counts shown. We find that there should be approximately (a) 55 galaxies at $z \sim 6$ at $J_{AB} \sim 27$, (b) 40 galaxies at $z \sim 7$ at $J_{AB} \sim 28$, and (c) 20 galaxies at $H_{AB} \sim 28$.

Densities at these levels point to a multiplex in the range of 20 to 40. Most notably, a minimum multiplex of ~20 would enable observations of all candidates with the largest photometric redshifts ($z > 7$) within the potential 38.5 arcmin$^2$ patrol field. Moreover, such a multiplex allows for an optimization of the target selection (by, e.g., morphology, OH-line avoidance, etc.) of galaxies with $1 < z < 4$ when the redshifts are known, or for systematic observations when the redshifts are not well constrained.

**2.2. Spatial Resolution**

The adopted requirements for spatial (angular) resolution and the necessary image quality (ensquared energy, EE) in high-redshift galaxies are the result of detailed simulations by Puech et al. (2008; 2010a; these proceedings). The starting point was to define an optimal scale over which galaxies begin to show discrete structure. Attempting to match an instrument to the scales of these features (rather than over-resolving them) then gives a sensitivity advantage in terms of surface brightness. Observations and theoretical models (e.g. Elmegreen & Elmegreen, 2005; Bournaud et al. 2007) reveal large structures of ~1 kpc scale. Translated to observables this requires good (20 to 30%) ensquared energies in the $H$-band over sampling scales of ~75 milliarcsec (mas); coarser sampling of ~100 to 150 mas is sufficient for recovery of large scale motions (Puech et al. 2008). In terms of the EE required from the AO system, the merger model shown in Fig. 3 indicates a minimum of ~25% EE in the $H$-band (within 75 mas, twice the pixel scale in the simulation) to reveal the different clumps in the emission-line maps, with greater EE leading to a clear enhancement.

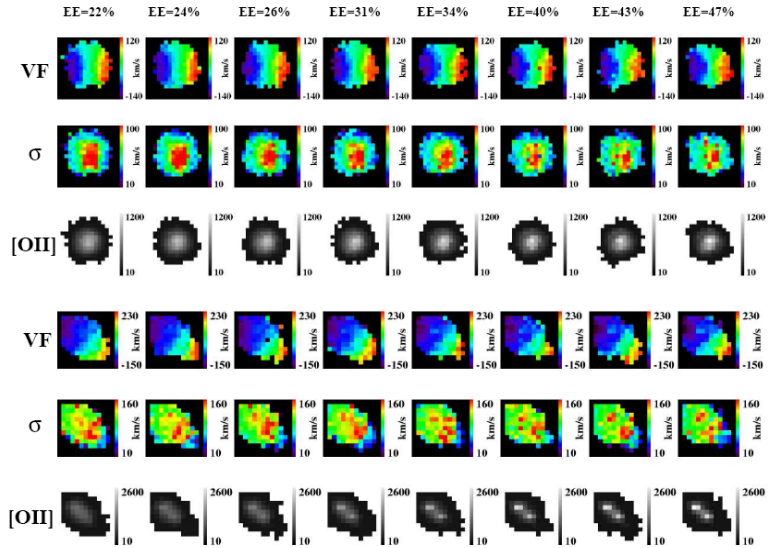

**Fig. 3:** Simulated $H$-band observations of a rotating disk galaxy (upper three panels) and a major merger (lower three panels) at $z = 4$ (Puech et al. 2008). The simulations are shown as a function of the ensquared energy (EE) in 75 mas pixels. Upper panels in each simulation: recovered velocity fields; middle panels: velocity dispersions; lower panels: emission-line maps.

The requirement for spatial resolution from the stellar populations case is harder to quantify. In some instances (e.g. dense stellar clusters, the innermost regions of external galaxies) there is a desire for the highest possible spatial performance/image quality. However, in terms of surveying large populations of resolved stars across whole galaxies, many of the targets are already known from deep *HST* imaging, i.e. the spatial resolution requirements are not exceptional, but the source magnitudes put them beyond the capabilities of 8-10m telescopes for absorption-line spectroscopy. There is a clear trade between survey speed and AO performance as the very best correction will, for now at least, only be possible over relatively small fields. At the other extreme, seeing-limited or ground-layer adaptive optics (GLAO) observations will not deliver sufficient contrast (particularly if considering AO correction at wavelengths shorter than 1 μm, e.g., Cunningham et al. 2008), so a capable wide-field AO system is clearly required.

The spatial resolution requirement adopted is good EE (≥30%) in 75 mas – a robust trade-off between AO performance, sensitivity, the different scientific inputs and cost. The EAGLE concept employs multi-object adaptive optics (MOAO) which uses laser and natural guide stars to tomographically map the atmospheric turbulence in a roughly cylindrical volume above the telescope. The corrections required for each target field within the focal plane are then calculated using this information, and introduced in each IFU channel using local deformable mirrors. Further details on the AO aspects of the study are provided by Rousset et al. (2010). MOAO is ideal for these science cases in that it operates over a large patrol field, with sufficient correction to study the critical sizes of small-scale galaxy structure.

**2.3. IFU Field-of-view**
This is primarily defined by the high-redshift cases where we require spatially-resolved spectroscopy, i.e. IFUs rather than slitlets or individual fibres. The field of each IFU needs to be well-matched to the targets which, from the isophotal sizes of galaxies observed with VLT-SINFONI (e.g. Lehnert et al. 2009), pushes for a field-of-view of ≥1.5×1.5 arcsec. We note that work to date has concentrated, necessarily, on the largest, highest surface-brightness galaxies. Typical M* galaxies at high redshifts will likely be somewhat smaller (e.g. the simulated M* merger at $z \sim 4$, Fig. 5), but maximising the IFU field-of-view retains the ability to also observe the larger systems. Moreover, IFUs of ≥1.5×1.5 arcsec would offer the potential to investigate multiple components of 'first light' systems. Many of the Lyman-break galaxies observed at $z \sim 5$ show evidence for multiple components over spatial extents of 1 to 2 arcsec (e.g. Douglas et al. 2009). If this trend continues to higher redshifts (as expected in a hierarchical mass-assembly model), spatially-resolved observations of each component will provide valuable insights into their overall properties (e.g., Fig 6).

The gain of IFU spectroscopy for the stellar case is huge compared to slitlet/fibre-based instruments. Crucially, in dense stellar fields, IFUs enable better subtraction of the local background. They also deliver simultaneous observations of multiple stars in a selected region of a galaxy (see, e.g., Fig. 7). For instance, observing a dozen or more stars in one region gives a much more efficient view of the population(s) than just observing one star at the same signal-to-noise ratio (SNR). We also note that IFUs with fields ≥ 1.5×1.5 arcsec are well-matched for other 'stellar' cases, such as spatially-resolved observations of massive star clusters in, e.g., the Antennae (cf. Bastian et al. 2006).

**2.4. Clustering/Tiling**
There are a number of cases which require relatively close clustering of IFUs, or even contiguous mapping of a region, e.g., mapping gas velocities in the regions near the Galactic Centre (Paumard et al. 2010). For the 'first light' case, in addition to follow-up of candidates from imaging, there is also the potential for blind spectroscopic surveys and/or observations near large galaxy clusters (which act as powerful gravitational lenses on the more distant systems lying behind, e.g. Swinbank et al. 2010). The lensed case provides a good example of the configuration requirements for the EAGLE IFUs: distributed, clustered, and contiguous mapping. For example, IFUs can be configured to target all of the giant arcs in a cluster in a single exposure, exploring lensed systems at $1 < z < 5$ (left-hand panel of Fig. 4). Similarly, IFUs can be used to observe the critical curves to search for galaxies at much greater redshifts ($7 < z < 20$), while the central part of the cluster can be observed with contiguous IFUs (right-hand panel of Fig. 4).

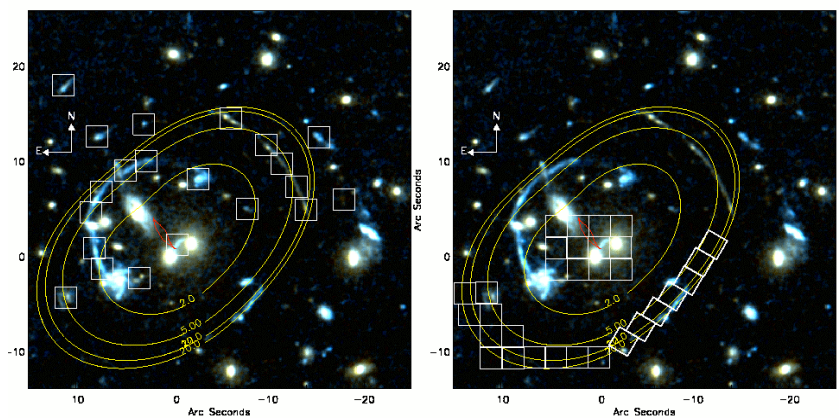

**Fig. 4:** *HST* image of the RCS0224-002 cluster ($z$ = 0.78). *Left-hand panel:* Example EAGLE configuration to observe giant arcs and arclets at $1 < z < 5$. *Right:* Mapping the critical curves ($z$ = 2 to 20), searching for highly magnified, very high-$z$ galaxies.

## 2.5. Spectral Resolving Power

A generic requirement is that the minimum spectral resolving power should be ~4000 so that the OH sky emission-lines can be subtracted effectively. Inclusion of a higher resolution mode is strongly desirable for quantitative stellar spectroscopy. Over the past decade the calcium triplet (CaT) has become a ubiquitous method to obtain estimates of metallicities and stellar kinematics (e.g. Tolstoy et al. 2001; Koch et al. 2007). The CaT comprises well-characterised lines from singly-ionized calcium (at 850, 854 and 866 nm). Results from VLT-FLAMES have demonstrated that, with careful calibration, [Fe/H] estimates from the CaT with low-resolution spectroscopy ($R$~6500) are in agreement with direct measurements from higher-resolution spectroscopy ($R$~20000) with the same instrument (Battaglia et al. 2008).

Selection of the desired resolving power is a trade-off between sensitivity and the precision with which one can recover robust abundance estimates. This trade-off is also strongly informed by the desire to obtain precise radial velocities. In preparation for the ESA *Gaia* mission, there has been significant effort toward modelling of the CaT region (e.g. Munari et al. 2000) as well as analysis of past data to understand the velocity precision obtained as a function of spectral resolving power and signal-to-noise (Munari et al. 2001). These results were employed to quantify the uncertainties on velocities from the Radial Velocity Experiment (RAVE; Steinmetz et al. 2006) finding:

$$\log (\Delta \text{RV}) = 0.6*(log\, \text{SNR})^2 - 2.4*log\, \text{SNR} - 1.75 log\, R + 9.36$$

where SNR is the signal-to-noise ratio of the observed CaT spectrum. This relation demonstrates the power of increasing $R$ compared to greater SNR in terms of the velocity precision achieved.

Radial velocities to precisions of better than 5 km s$^{-1}$ are required, preferably of order ±2-3 km s$^{-1}$. In Table 1 we summarise the spectral resolving power, as a function of SNR, required to obtain a velocity precision of 3 km s$^{-1}$. To push the observations as deep as possible, the typical SNR may be of order 10 to 15, arguing for greater $R$. We thus adopted $R = 10000$ as the baseline for the high-resolution mode. This specification is not only driven by the velocity precision for this case; inclusion of this higher spectral resolving power *significantly enhances the potential impact of EAGLE*, enabling a broad range of dynamical studies (stars and clusters, both resolved and unresolved) and atmospheric analysis of both young and old populations.

**Table 1: Effect of velocity precision (ΔRV) and signal-to-noise ratio (SNR) on the spectral resolving power (*R*) required.**

| ΔRV [km s$^{-1}$] | SNR | *R* |
|---|---|---|
| 3 | 20 | 7450 |
| 3 | 15 | 8650 |
| 3 | 10 | 11150 |

## 2.6. Wavelength coverage: Defining the blue cut-off for EAGLE

Strong motivation for coverage shortwards of 1μm is provided by the CaT. Observation of ~50nm centred on the CaT enables reasonable continuum normalization and subtraction of sky emission lines. Thus, the essential shortward cut-off for EAGLE is 0.84μm, with an optimum requirement of 0.80μm (delivering full coverage of the Paschen series, plus continuum for rectification). The potential for better AO performance at longer wavelengths has prompted efforts within the community to calibrate other potential diagnostics for abundance determinations (e.g. Davies et al., 2010). Research into novel near-IR diagnostics and their empirical calibration will no doubt continue over the coming decade, but it is hard to envisage a method supplanting the CaT as the *de facto* technique in the short-to-medium term (particularly given the strength of the CaT lines, meaning they can be used successfully in low metallicity regimes).

## 2.7. Target Acquisition Requirements

Many of the targets considered by the science cases will not be detectable in single acquisition images/individual IFU exposures. This puts relatively strict requirements on the positioning of each IFU within the field, so that repeat observations of faint targets (taken over several nights or interleaved with other programmes/calibrations) can be co-added accurately. The optimum requirement is that each IFU is correctly positioned to ±0.25 of a slice. The plate-scale at the GIFS is 3.838 mm/arcsec – assuming 37.5 mas slices, this corresponds to an accuracy of ±36 μm, which needs to include all error contributions from positioning and subsequent alignment. The key element here is not the absolute measurements but the repeatability of them. This requires that the EAGLE wavefront sensors relay sufficient information on the guide stars back to the telescope to maintain a stable plate scale in the GIFS focal plane.

## 2.8. Sky Coverage

Pushing the patrol field of EAGLE beyond the notional 5 arcmin diameter benefits the AO system by providing a larger area for selecting natural guide stars (NGS). For randomly selected fields at galactic latitudes of $50 < |b| < 90º$, the probability of five of more NGS with $R < 17$ is greater than 80% in the EAGLE patrol field (with an equivalent diameter of 7 arcmin). Rousset et al. (2010) provided results from the Besançon sky model for a slightly larger field (eqiv. to 7.3 arcmin diameter), with a probability of finding five NGS ($R < 17$) of >90% at $|b| = 60º$, and 60% at the Galactic Pole. These results are also supported by calculations by ESO (Calamida, 2009), whose results provide mean densities of 6.2 and 4.6 NGS ($R < 17$) at $|b| = 60$ and $90º$, respectively. Note that the 'Poor' NGS configuration used in the AO and performance simulations (see Table 2 for results) only has four NGS with $R < 17$, and still meets the image quality requirements (Rousset et al. 2010); i.e. the large EAGLE patrol field provides excellent sky coverage in terms of guide stars for AO, as well as maximising the number of science targets available.

## 2.9. Summary of Phase A Requirements

The baseline science requirements for the EAGLE Phase A study are summarised in Table 2.

**Table 2: Instrument requirements for EAGLE**

| Parameter | Requirement |
|---|---|
| Patrol field | ≥ 5' diameter |
| Science (IFU) field | ≥ 1.5" x 1.5" |
| Multiplex (No. science fields) | ≥ 20 |
| Spatial pixel scale | 37.5 milliarcseconds |
| Spatial resolution (in *H*-band) | Good ensquared energy (≥30%) in 2 x 2 spatial pixels |
| Spectral resolving power (*R*) | 4000 & 10000 |
| Wavelength coverage | *IZ*, *YJ*, *H*, *K* bands (0.8 – 2.45 μm) |
| Clustering/tiling | Distributed/clustered targets + ability for contiguous mapping |

## 3. PERFORMANCE

### 3.1. Simulations: Physics of high-redshift galaxies

Detailed simulations relating to the requirements and performances for the high-redshift galaxies case have, as noted earlier, been presented by Puech et al. (2008; 2010a), with details of EAGLE surveys discussed more specifically by Puech et al. (2010b). In Fig. 5 we show simulated major mergers from Puech et al. (2010a). Although using slightly larger spatial pixels than the EAGLE baseline, the simulations demonstrate the ability of the MOAO correction to recover spatially-resolved information in M* galaxies and even at lower masses, which is beyond the capabilities of the lower performance GLAO provided by the telescope (see Puech et al. 2010a for discussion).

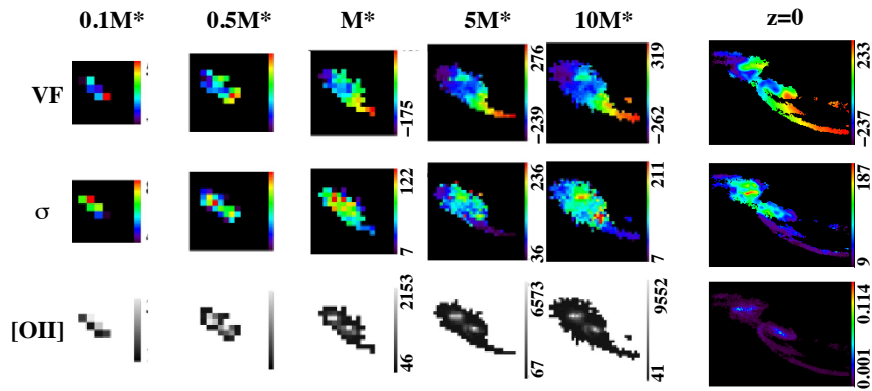

**Fig. 5:** Simulated observations of major mergers at $z = 4$ (50 mas pixels; Puech et al. 2010a), with the images corresponding to, from left to right, 0.4, 0.7, 0.9, 1.1 and 2.0 arcsec, depending on the galaxy mass. *Upper panels:* recovered velocity fields; *Middle panels:* velocity dispersions; *Lower panels:* emission-line maps. The input template ($z = 0$) is shown on the right.

## 3.2. Simulations: First light – characterising the most distant galaxies

In Fig. 6 we show simulated EAGLE observations of a clumpy system at very high redshift, using the morphology of a real (lensed) galaxy at $z = 4.9$ as the input template (left-hand panel in upper row; Swinbank et al. 2009). This is used as the source-plane input to the simulation ('input' image) with six clumps. The galaxy was shifted to an equivalent of $z = 7.15$, with a total magnitude of $J_{AB} = 27$, with the brightest clump with $J_{AB} = 28.4$. An integration time of 28 hours was used. The simulated $H$-band output image for one spectral element ($R = 4000$, with a spatial pixel scale of 37.5 mas) is shown in the middle panel. For comparison we also show the result of the same simulation with 5 mas sampling (corresponding to the diffraction limit of the E-ELT in the $K$-band), in which nothing is detected. Moreover, the chance of recovering any information on the low surface-brightness features between the bright clumps would be truly impossible. The lower panel of the figure shows the $J$-band spectra extracted from 37.5 mas spatial elements of the two brightest components, and the composite spectrum of the four brightest, in which absorption lines are recovered. Note that diffraction-limited sampling provides no useful data in such an example, as it effectively over-resolves the clumps, i.e., the EAGLE sampling is adequate to resolve the clumps of the object while preserving good sensitivity.

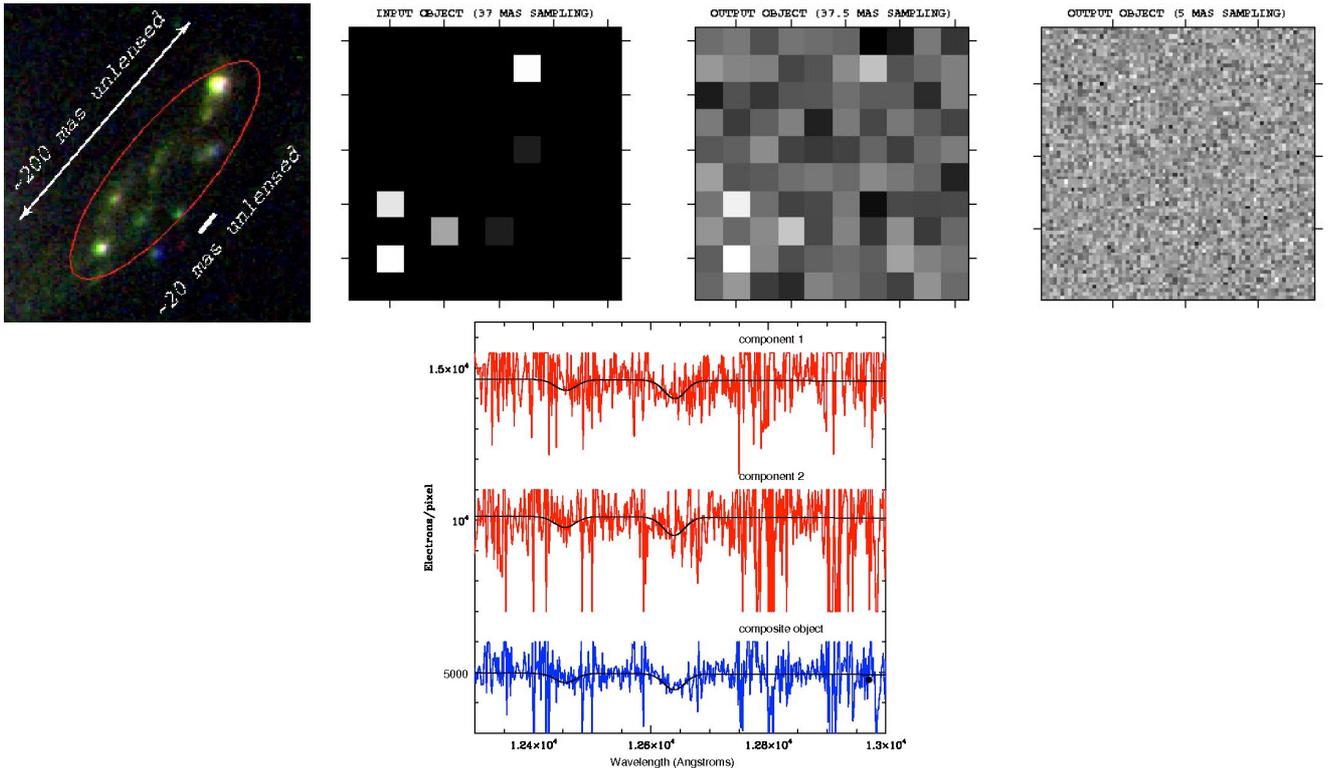

**Fig. 6:** Simulations of a clumpy high-$z$ galaxy. *Upper panel (from left to right):* Lensed image of a $z = 4.9$ galaxy (Swinbank et al. 2009), used an input for the clump morphology in the source plane; input image with six clumps at $z = 7.15$ (total magnitude: $J_{AB} = 27$; brightest clump: $J_{AB} = 28.4$); simulated EAGLE $H$-band cube for one spectral element at $R = 4000$ (between the OH lines) with a pixel scale of 37.5 mas and a total integration time of 28 hours; the result of the same simulation with spatial sampling matched to the E-ELT diffraction limit, which over-resolves the clumps. *Lower panel:* $J$-band spectra (37.5 mas sampling) of two brightest components, and composite of the four brightest.

## 3.3. Simulations: Resolved stellar populations in the Local Volume

A detailed discussion of the stellar populations case was given by Evans et al. (2010b), in which the EAGLE modelling tools were used to generate simulate CaT observations of evolved red giants. The magnitudes of the simulated stars were selected such that they were consistent with being members of the Sculptor Group galaxies (comprising two spirals at 1.9 Mpc and three at 3.6 to 3.9 Mpc), with some of the results summarised here in Table 2. The simulations investigated two sets of seeing conditions at a Paranal-like site; 0.9 arcsec seeing at a zenith distance (ZD) = 35º aims to provide an estimate of the performance in 'average' conditions that one might expect from the execution of a 'Large Programme'-like survey. The EAGLE performance depends on the relative positions and magnitudes of the guide stars within the field; here we employ MOAO simulations for two asterisms (see Rousset et al. 2010 for further detail). These simulations have also made use of existing *HST* observations of external galaxies, e.g., the *HST* Advanced Camera for Surveys (ACS) observations of NGC 300 (at 1.9 Mpc) as part of the ACS Nearby Galaxy Survey Treasury (ANGST; Dalcanton et al. 2009). The dense regions of such galaxies are very crowded yet the stars are resolved in the *HST* imaging, potentially yielding EAGLE spectroscopy of tens of stars per IFU (see Fig. 7).

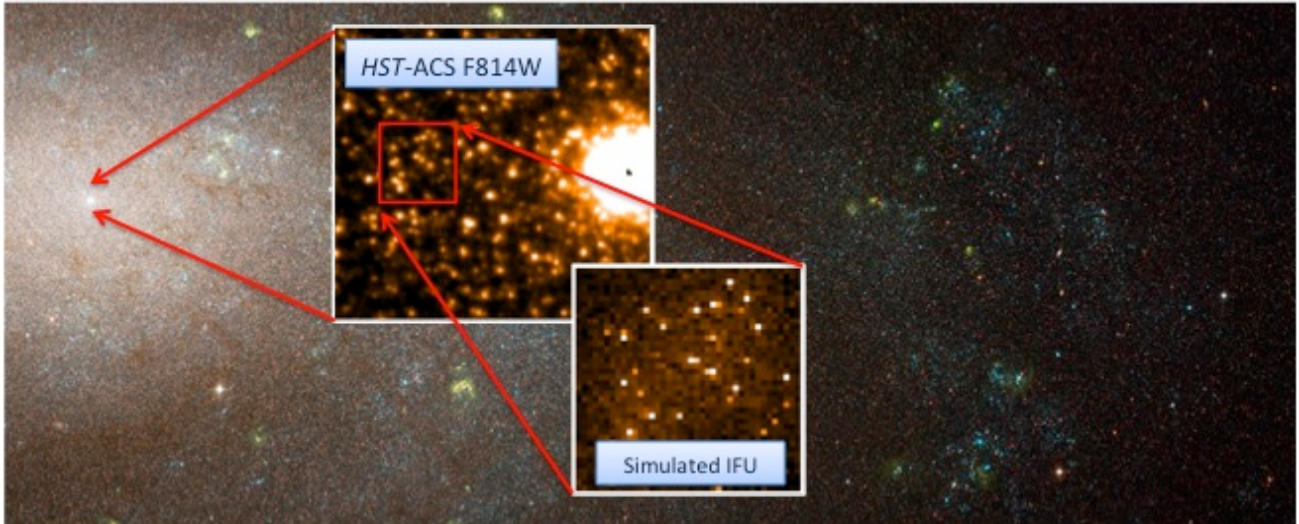

**Fig. 7:** *HST*-ACS three-colour image (F450W, F555W, F814W) of NGC 300 from the ANGST project (Dalcanton et al. 2009). The image spans approx 7.5 x 3 arcmin. The first inset is a ~6x6 arcsec zoom of the F814W image near the core. The second inset shows a simulated 1.5x1.5 arcsec EAGLE IFU field near the core, containing 31 stars with $22 < I_{VEGA} < 24.5$, illustrating the huge potential gain from the 'effective multiplex' in such a region .

The key result from the simulations is that a SNR ≥ 10 is recovered from a total integration of 10 hrs at $I_{VEGA}$ = 24.5, in mean seeing, in both configurations of guide stars, at $R$ = 10000. This is four magnitudes deeper than FLAMES-Giraffe at the VLT using the LR08 grating (which includes the CaT, at $R$ = 6500) with the same exposure time. Observations with $R$ = 4000 can obviously push even deeper (see Evans et al. 2010b).

**Table 2:** Continuum signal-to-noise ratio (SNR) obtained from simulated EAGLE observations of the CaT ($R$ = 10000, total integration time = 10 hrs) for two sets of conditions at a Paranal-like site and two configurations of natural guide stars (NGS).

| $I_{VEGA}$ | Seeing: 0.9" at ZD=35° | | Seeing: 0.65" at ZD=0° | |
| --- | --- | --- | --- | --- |
| | 'Good' NGS | 'Poor' NGS | 'Good' NGS | 'Poor' NGS |
| 22.5 | 40 | 27 | 56 | 48 |
| 23.5 | 16 | 11 | 28 | 24 |
| 24.5 | 8 | 4 | 13 | 10 |

Note that the *I*-band MOAO point-spread functions (for mean Paranal seeing of 0.65 arcsec; Rousset et al. 2010) deliver an *encircled* energy which is comparable to that delivered by the wide-field channel of the *HST* Advanced Camera for Surveys (ACS; Sirianni et al. 1995) – the ability to combine this excellent image quality over a 38.5 arcmin$^2$ field-of-view, while harnessing the sensitivity of the 42 m primary of the E-ELT, is EAGLE's unique capability in this case.

### 3.4. Overall performance
Between the OH lines and atmospheric absorption windows (approx. two thirds of the 0.8 to 2.5μm range), EAGLE will offer the best near-IR spectroscopic performance (in terms of survey speed) when compared to *JWST*-NIRSPEC and other spectrographs proposed for the ELTs. The continuum between the OH lines as seen from the ground is only a factor of two or three greater than that seen from space, enabling the E-ELT to deliver better spectroscopic performance than that of the *JWST*. The fine spatial resolution achieved using MOAO will provide EAGLE with an optimum point-source sensitivity, without critical sampling of the diffraction limit. In Fig. 8 we show a comparison of the performance of MOAO compared to GLAO and multi-conjugate adaptive optics (MCAO), with the latter calculated for the Thirty Meter Telescope (TMT) by way of a further comparison. MOAO outperforms both GLAO and TMT-MCAO sensitivities. It will also outperform laser-tomography adaptive optics (LTAO) in terms of the field-of-view available for targets, in addition to the fact that if the pixel scale is too fine it would result in readout-noise limited observations and thus reduced performance.

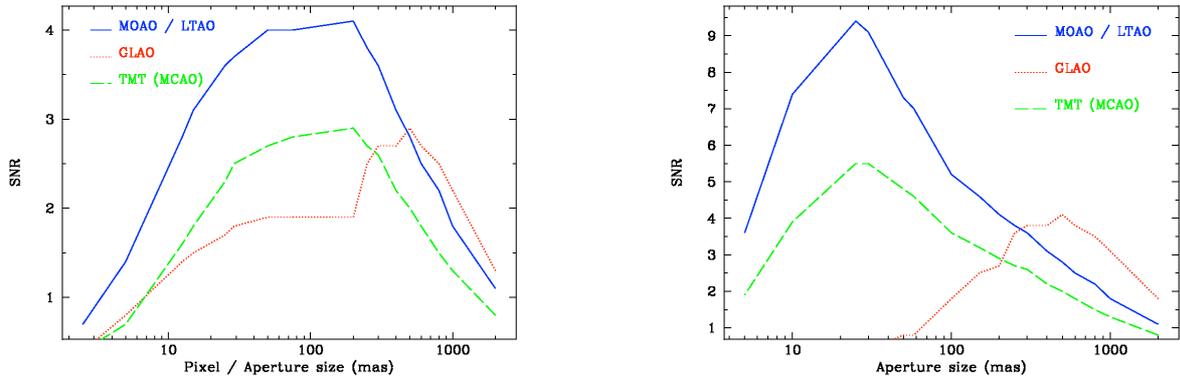

**Fig. 8:** Signal-to-noise ratio (SNR) vs. 'slit' aperture (pixel size) for $J_{AB} = 27$. *Left:* A 200 mas extended source with the SNR calculated for a matched aperture, i.e. ≥ 200 mas. For pixels < ~50 mas the performance is readout-noise limited and the SNR decreases despite the constant (200mas) aperture. *Right:* A point source, with the SNR calculated over the aperture indicated.

## 4. SYNERGIES

EAGLE will have strong synergies with both the *JWST* and ALMA. Others facilities, such as the VLTs and VISTA, will also provide support in the form of pre-imaging and target selection. The *JWST* has superb imaging sensitivity and so offers a vast discovery space, but is limited in terms of its spectroscopic sensitivity due to its (relatively) small collecting area – EAGLE will provide the capability to obtain spectroscopy of the bulk of the targets discovered/identified with the *JWST*. Note that the pixel scales between *JWST*-NIRCam (32 mas) and EAGLE (37.5 mas) are well matched, delivering spectroscopic follow-up at comparable angular resolution. The *JWST* deep fields will also cover areas which are perfectly matched to the EAGLE patrol field. In essence, the *JWST* will provide initial images and spectral energy distributions, EAGLE will reveal the astrophysics behind these discoveries.

ALMA will have a significant headstart in time over the E-ELT but EAGLE, with its high multiplex, can quickly catch-up. Together they will reveal the star-formation rates and kinematics of distant galaxies at comparable resolutions. ALMA will probe the molecular gas, dust, and the ionized species of atomic and simple molecules, while EAGLE will probe the warm and hot ionized medium in the ISM and the stellar continuum – i.e. providing very complementary observations of different gas phases, in both nearby and distant galaxies. This synergy is neatly illustrated by recent observations of individual star-forming clumps at $z = 2.3$ (Swinbank et al. 2010) – this is currently only possible in lensed systems, but near-infrared observations on the E-ELT will provide unique views of star-formation in distant galaxies. The synergy of the E-ELT with ALMA is potentially strong, and will be best realised by EAGLE.

Further into the future, one can foresee valuable observing programmes using the Square Kilometre Array (SKA) to measure H I 21 cm gas properties of galaxies with extant E-ELT and ALMA data, and also combinations of the above with the *International X-ray Observatory*.

## 5. SUMMARY

The capabilities of EAGLE, combined with the sensitivity of the E-ELT, will bring a large, detailed survey of high-redshift galaxies within our reach. The fine spatial resolution (75 milliarcsec) will allow us to study the physical processes taking place at sub-kpc scales, providing the perfect complement to studies of cold gas and dust with ALMA at approximately the same resolution. With greater spectroscopic sensitivity than the *JWST*, with a good multiplex, and with spatially-resolved mapping from the IFUs, EAGLE will deliver a definitive view of galaxy evolution in only a few hundred hours of observations – i.e. a well targeted ESO Large Programme.

Similarly, the combination of sensitivity, spectral coverage, wide field and multiplex will provide observations of galaxies at $7 < z < 10$ in the way that we now almost take for granted at $3 < z < 5$ with 8 to 10 m class telescopes. In a few hundred hours of observations, EAGLE will study the spatially-resolved properties of ~100 Lyman-break galaxies in the early Universe in exquisite detail, providing the first insights into the dynamics and characteristics of the ISM in these elusive galaxies. Such observations are at the very forefront of astrophysics, building on the capabilities and potential of the *JWST* and ALMA.

EAGLE will also provide a unique view of the stellar populations in external galaxies, with observations in their extended outer halos and toward core or bulge regions at the same time – an unmatched capability. The area covered by individual IFUs enhances the multiplex by allowing us to study multiple stars within each channel, as well as the fainter unresolved population. This 'effective multiplex' is a huge advantage. A large sample (>>1000) of stars will be observable in galaxies beyond the Local Group in 200-300 hrs. Remembering this is absorption-line spectroscopy ($R$~10000), this is well beyond the capability of current instrumentation, and any other proposed E-ELT instrument.

EAGLE will be a general facility-class instrument, which will serve a broad community of users. It uses the E-ELT at its best, over a wide field-of-view and with AO-corrected images, efficiently tackling some of the most prominent science cases advanced for building the observatory.

*Acknowledgements:* Thanks to Thibaut Paumard and Yann Clenet (LESIA, Obs. Paris), Yanbin Yang (GEPI, Obs. Paris) and Dave Alexander (Durham) for their inputs to the EAGLE science case. This work was funded by the UK Science and Technology Facilities Council (STFC), the European Southern Observatory (ESO), the Centre National de la Recherche Scientifique (CNRS), the Agence Nationale de la Recherche (contract ANR-06-BLAN-0191) and the European FP7 Infrastructure programme (contract 211257).

# REFERENCES

Bastian et al. 2006, A&A, 445, 471
Battaglia et al. 2008, MNRAS, 383, 183
Bournaud, Elmegreen & Elmegreen, 2007, ApJ, 670, 237
Bouwens et al. 2008, ApJ, 686, 230
Bouwens et al. 2010, ApJ, 709, L133
Calamida, 2009, ESO, E-TRE-ESO-080-0588 Issue 1
Cunningham et al. 2008, SPIE, 6986, 20
Dalcanton et al. 2009, ApJS, 183, 67
Davies et al. 2010, MNRAS, arXiv:1005.1008
Douglas et al. 2009, MNRAS, 400, 561
Elmegreen & Elmegreen, 2005, ApJ, 627, 632
Evans et al. 2010a, A&G, 51b, 17
Evans et al. 2010b, AO4ELT conf., 1004 ; arXiv :0909.1748
Flores et al. 2006, A&A, 455, 107
Förster Schreiber et al. 2004, ApJ, 616, 40
Förster Schreiber et al. 2006, ApJ, 645, 1062
Förster Schreiber et al. 2009, ApJ, 706, 1364
Koch et al. 2007, AJ, 133, 270
Lehnert et al. 2009, ApJ, 699, 1660
Marchesini et al. 2007, ApJ, 656, 42
McLure et al. 2010, MNRAS, 403, 960
Munari & Castelli, 2000, A&AS, 141, 141
Munari et al. 2001, BaltA, 10, 613
Oesch et al. 2009, ApJ, 690, 1350
Oesch et al. 2010, ApJ, 709, L16
Paumard et al. 2010, AO4ELT conf., 1003
Puech et al. 2006, A&A, 455, 131
Puech et al. 2008, MNRAS, 390, 1089
Puech et al. 2010a, MNRAS, 402, 903
Puech et al. 2010b, AO4ELT conf., 1005; arXiv:0909.1747
Reddy et al. 2008, ApJS, 175, 48
Rousset et al. 2010, AO4ELT conf., 2008; arXiv:1002.2077
Scoville et al. 2007, ApJS, 172, 1
Sirianni et al. 1995, PASP, 117, 1049
Steinmetz,et al. 2006, AJ, 132, 1645
Swinbank et al. 2009, MNRAS, 400, 1121
Swinbank et al. 2010, Nature, 464, 733
Tolstoy et al. 2001, MNRAS, 327, 918
Yang et al. 2008, A&A, 477, 789